\documentclass[12pt]{iopart}
\usepackage{iopams}  
\usepackage{epsfig,graphicx}                         %
\usepackage{natbib}                                                           %
\usepackage{picins}

\bibpunct[,]{(}{)}{;}{a}{}{,}

\def\pderiv(#1/#2){\frac{\partial#1}{\partial#2}}                   %

\begin{document}

\title [Event rate of EMRIs and cusp re-growth]
       {The impact of realistic models of mass segregation 
        on the event rate of extreme-mass ratio inspirals and 
        cusp re-growth}
      
\author{Pau Amaro-Seoane$^{1,\,2}$ and Miguel Preto$^{3}$}

\address{
$^1$Max Planck Institute for Gravitational Physics\\
                 (Albert Einstein Institute), Potsdam-Golm, Germany\\
$^2$Institut de Ci{\`e}ncies de l'Espai (CSIC-IEEC)\\
                 Campus UAB, Torre C-5, parells, $2^{\rm na}$ planta, 
                 ES-08193, Bellaterra, Spain\\
$^3$Astronomisches Rechen-Institut (ZAH)\\
                 M{\"o}nchhofstr. 12-14
                 D-69120 Heidelberg, Germany
}

\ead{Pau.Amaro-Seoane@aei.mpg.de,\,miguelp@ari.uni-heidelberg.de}

\begin{abstract}
One of the most interesting sources of gravitational waves (GWs) for {\em LISA}
is the inspiral of compact objects on to a massive black hole (MBH), commonly
referred to as an ``extreme-mass ratio inspiral'' (EMRI). The small object,
typically a stellar black hole (bh), emits significant amounts of GW along each
orbit in the detector bandwidth.  The slowly, adiabatic inspiral of
these sources will allow us to map space-time around MBHs in detail, as well as
to test our current conception of gravitation in the strong regime.
The event rate of this kind of source has been addressed many times in the
literature and the numbers reported fluctuate by orders of magnitude. 
On the other  hand, recent observations of the Galactic center revealed a 
dearth of giant stars inside the inner parsec relative to the numbers 
theoretically expected for a fully relaxed stellar cusp. The possibility of 
unrelaxed nuclei (or, equivalently, with no or only a very shallow cusp, or core) 
adds substantial uncertainty to the estimates. Having this timely question in mind, 
we run a significant number of direct-summation $N-$body simulations with up 
to half a million particles to calibrate a much faster orbit-averaged Fokker-Planck 
code. We show that, under quite generic initial conditions, the time required for 
the growth of a relaxed, mass segregated stellar cusp is shorter than a Hubble time 
for MBHs with $M_\bullet \lesssim 5 \times 10^6 M_\odot$ (i.e. nuclei in the range of 
{\em LISA}). We then investigate the regime of strong mass segregation (SMS) for 
models with two different stellar mass components. Given the most recent stellar
mass normalization for the inner parsec of the Galactic center,  SMS has the 
significant impact of boosting the EMRI rates by a factor of $\sim 10$ in 
comparison to what would result from a $7/4-$Bahcall \& Wolf cusp resulting in
$\sim 250$ events per Gyr per Milky Way type galaxy. Such intrinsic rate should
translate roughly into $\sim 10^2-7 \times 10^2$ sbh's EMRIs detected by LISA 
over a mission lifetime of $2$ or $5$ years, respectively), depending on the 
detailed assumptions regarding LISA detection capabilities.    
\end{abstract}

\maketitle

\section{Introduction}

Nowadays it is well-established that a massive dark object, very possibly a massive 
black hole (MBH) with a mass of 
about $4 \times 10^6 M_\odot$, is lurking in the centre of the Milky Way \citep{
EisenhauerEtAl05,GhezEtAl05,GhezEtAl08,GillessenEtAl09}. While there is an emerging 
consensus about the origin and growth of supermassive black holes (SMBH, with masses 
about or larger than $10^8 M_\odot$) \citep{2005SSRv..116..523F,2008ApJ...676...33D,
2010A&ARv..18..279V}, MBHs with smaller masses such as the one in the Galactic centre 
remain a (relatively) understudied enigma. One of the keys to understanding the growth 
and evolution of MBHs in this lower mass range 
resides in the dynamics of stars in their vicinity. This is the case mainly because 
relaxation times there are low enough that the surrounding stellar systems should have had
enough time---through two-body relaxation {\em alone}---to evolve towards a steady-state
which is independent of the particular initial conditions at the time of formation. The
Galactic center is thought to fulfill such condition. It is the universality of such
relaxed stellar nuclei that gives us a crucial predictive power on the expected 
properties of the MBH environment, on the stellar candidates for close interaction
with the central MBH and on the resulting gravitational wave (GW) signatures. If, on the
contrary, non-relaxed systems were generic, then one would need to resort to {\em case-by-case
modelling of each galactic nucleus}.

The ideal probe for these innermost regions of galaxies is the GW radiation that is emitted 
by stellar bhs and other compact objects that come very close to the MBH. One of the main 
channels for interaction between stars and a central MBH is the adiabatic, slow inspiral of 
compact remnants (CR) into the MBH due to the emission of GWs---an EMRI.  During such an event,
the small body effectively acts as a probe of spacetime close to the MBH as its orbit slowly 
shrinks due to the energy and angular momentum lost in the form of GW radiation. In case of
$10^5-10^6 M_\odot$ MBHs,  after some $\sim 10^4-10^5$ orbits  in the LISA band ($f_{\rm orb} 
\gtrsim 10^{-4}$ Hz and a periapsis $a \lesssim \rm{\rm few} \times R_{\rm Schw}$, since we only 
consider sources which are completely embedded in the band, and not bursting sources), the 
small body eventually merges with the MBH. The information contained in the waves will allow 
us to determine the parameters of these binary system with an unprecedented accuracy 
\citep[see for instance][]{2004PhRvD..69h2005B,BabakEtAl10}, corroborate the existence of 
MBHs and maybe even provide the first direct detection of an intermediate MBH (in case the 
primary is $\sim 10^{3-4} M_\odot$). 

{\em LISA} will thus scrutinize exactly the mass range about which electromagnetic observational
information is currently lacking. In its most general form, the EMRI problem---the astrophysical 
modelling of event rates and parameters for EMRIs---spans many orders of magnitude.
From the bulge regions at few$\times 10$ pc, where the dynamics is essentially collisionless
--but from where single stellar bhs and binaries with CRs originate;  down to the parsec scale 
of the nucleus itself which evolves secularly over (local) relaxation timescales; and then further down 
to milliparsec scales where relativistic effects start to dominate the evolution. But,
however, once a steady state configuration establishes itself in the central parsec region, the EMRI
rates are rather expected to depend strongly on the (universal) density distribution of CRs within 
(in order of magnitude) ${\cal O}(0.01 \rm{pc})$ from the hole. This is indeed the region from which these 
inspiralling sources are expected to originate \citep{HopmanAlexander05}. The dynamics in this tiny 
volume has been rather unexplored until the relevance of EMRIs and sub-parsec observations of the 
Galactic center have raised its interest. Since then, many authors have devoted a number of works to 
the analysis of this peculiar regime \citep{SR97,Freitag03,AH03,HopmanAlexander06}.

We discuss in this work the stellar distribution of dense stellar systems around MBHs in the
LISA mass range. Realistic modeling of mass segregation---which is the natural outcome for any 
realistic stellar population---will strongly impact the expected EMRI rates, since it favors 
the accumulation of heavier objects towards the center \citep{HopmanAlexander06,2009ApJ...697.1861A,
PA10}. In Section $2$, we begin by summarizing the results obtained by \cite{PA10} that show how
to calibrate the FP calculations with direct $N$-body simulations\footnote{Direct $N$-body simulations
compute the gravitational accelerations between particles using the exact Newton's law, without
introducing any approximations to compute the gravitational potential \citep{2008gady.book.....B}.}; 
then, still in the same section, we present new results concerning the robustness of $N$-body 
realisations of stellar cusp growth with respect to the total particle number $N$. In section $3$, 
we present new results on the growth of stellar cusps from a variety of initial conditions resulting 
from carving a cavity in the star's phase space distribution function. This is motivated from a variety 
of astrophysical mechanisms that may lead to cusp destruction; and these mechanisms are critically 
assessed in the end of the section. We show that, 
under very generic circumstances, the time required for the growth of a cusp is shorter than a 
Hubble time. Therefore, quasi-steady, mass segregated, stellar cusps are expected to be common 
around MBHs in the LISA mass range. But see \cite{2010ApJ...718..739M} and \cite{2010arXiv1010.1535M} 
for different perspectives. EMRI detection rates for {\em LISA} are expected to peak for $M_\bullet 
\sim 10^5-10^6 M_\odot$ \citep{Gair09} leading us to conclude that at least a sizeable fraction of 
these events should originate from strongly segregated cusps. Finally, in Section $4$ we present 
new estimates on the expected EMRI rates in mass segregated nuclei and conclude that our realistic 
modeling of mass segregation has a significant impact on these rates.

\section{Mass segregation}

The distribution of stars around a massive black hole is a
classical problem in stellar dynamics \citep{1976ApJ...209..214B,1977ApJ...211..244L}.
\cite{1976ApJ...209..214B} have shown, through a kinetic treatment that, within the radius
of gravitational influence of the hole $r_h$, in case all stars are of the same mass, this 
quasi-steady distribution takes the form of power laws, $\rho(r) \sim r^{-\gamma}$, in physical 
space with $\rho(r)$ the stellar density at a radius $r$ and $f(E) \sim E^p$ in energy space 
(with $E$ the energy and $\gamma=7/4$ and $p=\gamma-3/2=1/4$)\footnote{We note that 12 years 
{\em before} the work of BW, \cite{Gurevich64} derived a similar solution for how electrons 
distribute around a positively charged Coulomb center, which is the equivalent of the MBH in 
our case.}. This is the so-called {\it zero-flow solution} for which the net flux of stars 
in energy space is {\em precisely} zero. \cite{2004ApJ...613L.109P} and \cite{2004ApJ...613.1133B} 
were the first to demonstrate the robustness of the corresponding direct-summation 
$N$-body realizations, and have therefore validated the assumptions inherent to the 
Fokker-Planck (FP) approximation---namely, that scattering is dominated by uncorrelated, $2$-body 
encounters  and, in particular, dense stellar cusps\footnote{In this work, a nucleus is said
to be a {\it core} if $\gamma<1$; it is said to be a {\it cusp} if $\gamma>1$.}
populated with stars of the {\it same mass} 
are robust against ejection of stars from the cusp. The latter result is not trivial as for a BW 
$\gamma=7/4$ cusp stellar densities are extremely high at the center and the fraction of stars with 
speeds close to the escape velocity from the cusp is quite high at all radii $r \lesssim r_h$, with 
$r_h$ the influence radius of the MBH \citep{2010GWN.....3....3P}.

Single mass models are very poor approximations of real stellar populations. To 
first order of approximation, an evolved stellar population can be represented by two (well-separated)
mass scales: one in the range ${\cal O}(1 M_\odot)$ corresponding to low mass main-sequence stars, 
white dwarfs (WDs) and neutron stars (NSs); another with ${\cal O}(10 M_\odot)$ representing stellar
bhs. Therefore, for simplicity, here we restrict our discussion to models with two mass components and
leave the more general case to another work in preparation \citep{PA-2-10}.

When stars of two different masses are present, there is mass segregation which is a process by
which the heavy stars accumulate near the center while the lighter ones float outward
\citep{1987degc.book.....S,2007MNRAS...374.703K}. Accordingly, stars with different mass get 
distributed with different density profiles. \cite{1977ApJ...216..883B}, henceforth BW77, have 
argued heuristically that a scaling relation $p_i = m_i/m_j \times p_j$ (where the subindices $i$, 
$j$ refer to the light or heavy components) establishes itself and depends only on the mass ratio. 
Here, as in the single-mass case, the crucial assumption is that all components  are abundant 
enough that they undergo enough scattering among themselves and with the other components as to 
stabilize into an approximate {\it zero-flow} solution.  Obviously, this cannot happen independently 
of the number fraction of the different stellar masses \citep{2009ApJ...697.1861A,PA10}.  In the 
realistic situation where the number fraction of heavy objects (in our case, stellar bhs) is small, 
a new solution coined by \cite{2009ApJ...697.1861A} as {\it strong mass segregation} (SMS) obtains 
with density of heavy objects scaling as $\rho_H(r) \sim r^{-\alpha}$, where $\alpha \gtrsim 2$. The 
solution has two branches and can be parametrized by the parameter

\begin{equation}
\Delta = \frac{D_{HH}^{(1)}
+D_{HH}^{(2) }}{D_{LH}^{(1)} +D_{LH}^{(2)}} \approx \frac{N_H m_H^2}{N_Lm_L^2}
\frac{4}{3+m_H/m_L},
\end{equation}

\noindent
where $N_L$ and $N_H$ are the total number of light and heavy stars, $m_L$ and 
$m_H$ are the corresponding individual masses. $\Delta$ provides a measure of the
importance of the heavy star's self-coupling relative to the light-heavy
coupling (in terms of the $1^{\rm{st}}$ and $2^{\rm{nd}}$ order diffusion
coefficients); and it depends essentially on the mass {\it and} number ratios,
which is one parameter more than proposed by BW77. The $weak$  branch, for
$\Delta > 1$ corresponds to the scaling relations found by BW77; while the
$strong$ branch, for $\Delta < 1$, generalizes the BW77 solution\footnote{The 
choice of the names is based upon the resulting slopes in the density profiles, 
which are steeper ({\em stronger}) or shallower ({\em weaker})}. Stellar 
populations with continuous star formation and an initial mass function (IMF) 
given by $dN/dM \propto M^{-\alpha}$ will be characterized by $\Delta < 1$ if $\alpha 
\gtrsim 1.8$ and $\Delta < 1$ otherwise; and, in particular, Salpeter and Kroupa's 
IMF generate evolved stellar populations with $\Delta <1$ \citep{2009ApJ...697.1861A}.

{\bf Figure 1.} Evolution of density profiles. Mass density profiles, $\rho_L(r)$ 
(left panels) and $\rho_H(r)$ (right panels) at the end of the integrations, after 
$\approx 0.2 T_{\rm{rlx}}(r_{\rm h})$. Red curves are from FP calculations, green
\parpic[r][r]{\mbox{\resizebox{0.4\hsize}{!}
        {\includegraphics{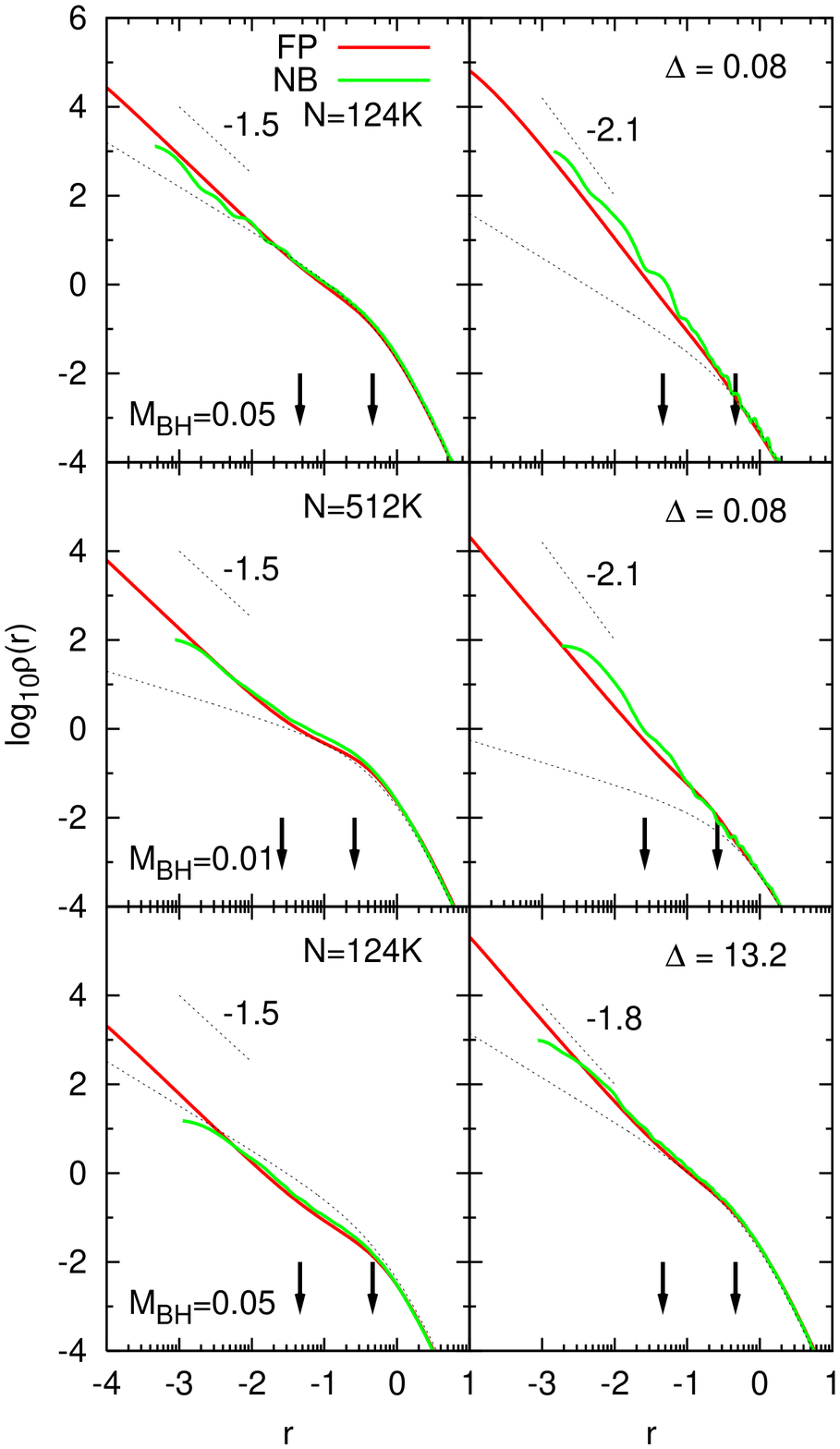}}}}
\vspace{-4pt}
\noindent
curves are from NB simulations. The agreement between both methods is quite
good. The mass ratio between heavy and light stars is $R=10$, representing
the expected typical mass ratio between light stars (MS stars, WDs and NSs)
and stellar bhs, as explained in the text; the number
fraction of heavy stars $f_H = 2.5\times 10^{-3}$ (top and middle panels) and
$f_H=0.429$ (lower panels), corresponding to the strong and weak segregation
regimes  respectively. The initial condition is a Dehnen profile with central
slope $\gamma=1$ for the top and bottom panels \citep{PA10}, $\gamma=1/2$ in 
the middle panel; a central MBH with $5\%$ of the total mass of the cluster 
and $1\%$ likewise. The particle number is $N=124,000$ (top and bottom) and 
$N=512,000$ (middle). The asymptotic slope $\gamma_H$ decreases from $\gtrsim 2$ 
to $\approx 7/4$ when moving from the strong to the weak branch of the solution.
The asymptotic slope $\gamma_H \approx 3/2$ throughout, or just slightly below
this value. The arrows point to radii $r_h$ and $0.1 r_h$.
\vspace{8pt}

There is a straightforward physical interpretation for the strong branch of
mass segregation.   In the limit where heavy stars are very scarce, they barely
interact with each other and instead sink to the center due to dynamical
friction against  the sea of light stars. Therefore, a quasi-steady state
develops in which the heavy star's current is {\em not} nearly zero and thus 
the BW77 solution does not hold exactly anymore. Indeed, in the limit where the 
number fraction $f_H$ of heavy stars is vanishingly small, as is the case of 
nuclei with realistic IMFs, the stellar potential is dominated by the light 
component. In this case, the light stars should evolve as if in isolation and 
develop a $\gamma_L \sim 7/4$ density cusp. The scarce heavy stars sink to the 
center  due to dynamical friction against the background of light stars, and 
will not exert any significant back-reaction on them \citep{2010GWN.....3....3P}.

Figure 1 displays the FP and NB evolutions of the spatial density $\rho_L(r)$ and 
$\rho_H(r)$ for models with two mass components corresponding to different initial 
profiles, MBH masses and total particle number $N$. The starting models are either 
$\gamma=1$ or $\gamma=1/2$ Dehnen profiles for both components with a MBH of $1\%$ 
or $5\%$ of the total mass of the cluster. Dehnen density profiles are defined by
$\rho(r)=(3-\gamma)M_{TOT}/4\pi r^\gamma (r_b+r)^{4-\gamma}$, have total stellar mass 
$M_{TOT}$, an inner (outer) logarithmic slope $\gamma$ ($4-\gamma$) and a break radius 
$r_b=1$ (which is larger than $r_h$ in all cases). We adopt units where $G=M_{TOT}=1$. The 
density of both components reaches a quasi-steady state within $\sim 0.2 T_{\rm rlx}(r_h)$, 
where $T_{\rm rlx}(r_h)$ relaxation time measured at the influence radius \citep{PA10}. 
The top and middle panels display the strong mass segregation solution with $\gamma_H 
\sim 2.1$ as expected for $\Delta=0.08$ ($f_H=2.5 \times 10^{-3}$); while, in the bottom 
panel, $\Delta=13.2$ ($f_H=0.429$) displays the weak solution for which $\gamma_H \sim 
7/4$. The former value was chosen to be close to the number fraction of stellar bhs
to be expected from a standard Salpeter or Kroupa's IMF; the latter value is chosen
to be representative of the regime of weak segregation studied by BW77. One can see 
from Figure 1 that in the case of weak segregation $\rho_H > \rho_L$ everywhere due 
to the extremely high number of heavy objects; in contrast, in the SMS regime $\rho_H 
> \rho_L$ only for $r \lesssim 0.01 r_h$ (and the light objects dominate in number 
almost everywhere). In all cases the asymptotic slopes are valid within $\sim 0.1 
r_h$ and are totally predictable once $\Delta$ is known. These results agree pretty
well with the predictions for the SMS regime \citep{2009ApJ...697.1861A}.

The particle number in our direct-summation $N-$body simulations sample ranges from 
$N=124,000$ to $N=512,000$; our results do not show evidence of any dependence on total 
$N$, nor on the initial value of $\gamma$, once the results are re-scaled appropriately 
({\it i.e.} measured in terms of the relaxation time). The agreement between NB and FP 
methods is quite good in all cases.

\section{Cusp Re-growth}

\subsection{Current observations: A missing cusp}

We have seen that theory predicts a steady state cusp that reaches extremely high densities in
the center near the MBH. Furthermore, given a normalization at, say, $r_h$ and a knowledge
of the stellar mass function (and thus of $\Delta$), the density profile inside $r_h$ becomes 
completely determined. But observations are much more complicated to interpret. First, one 
must realize that there are very few galaxies for which the influence radius $r_h$ can be 
resolved. In fact, except for the nearest galaxies, $r_h$ covers an angular region in the sky 
which is too small to be resolved even with the HST. The HST has an angular resolution 
of $\sim 0.1''$. In the case of galaxies in the Virgo cluster, at $\sim 17$ Mpc of distance,
it can only resolve regions with linear sizes $\gtrsim 8.25$ pc. Therefore, HST would not
be able to resolve SgrA*'s radius of influence if it were at the distance of Virgo. Since
$r_h \propto M_\bullet^{1/2}$, it can only start to resolve the influence radius of Virgo's MBHs 
that have masses $M_\bullet \gtrsim 4 \times 10^7 M_\odot$. Second, even in the few cases for 
which $r_h$ can be resolved to some extent, it still is necessary to assess whether the 
observed stars (only those that are bright enough to be detected) really trace the 
underlying (dynamically dominant) invisible population. Third, given the fact that, as we 
have seen, stars tend to segregate by mass, there is an extra uncertainty related to the 
unknown stellar mass function. Moreover, there 
are indications that star formation events are common in galactic nuclei and furthermore 
that the resulting IMF in these sub-parsec regions may be substantially different from 
that of the field stars and biased towards heavy masses \citep{2010ApJ...708..834B}. 
Finally, it is necessary to deproject the observations and, in the (inevitable) absence 
of complete knowledge of phase space coordinates, one must rely on kinematic assumptions 
regarding the (an-)isotropy of stellar velocities and on the three dimensional 
shape of the stellar system.

\vspace{8pt}
{\bf Figure 2.} Time for cusp re-growth. Single-mass relaxation time at $r_h$
for single-mass cored models as a function of MBH mass. The shaded area covers
\parpic[r][r]{\mbox{\resizebox{0.55\hsize}{!}
        {\includegraphics[angle=270]{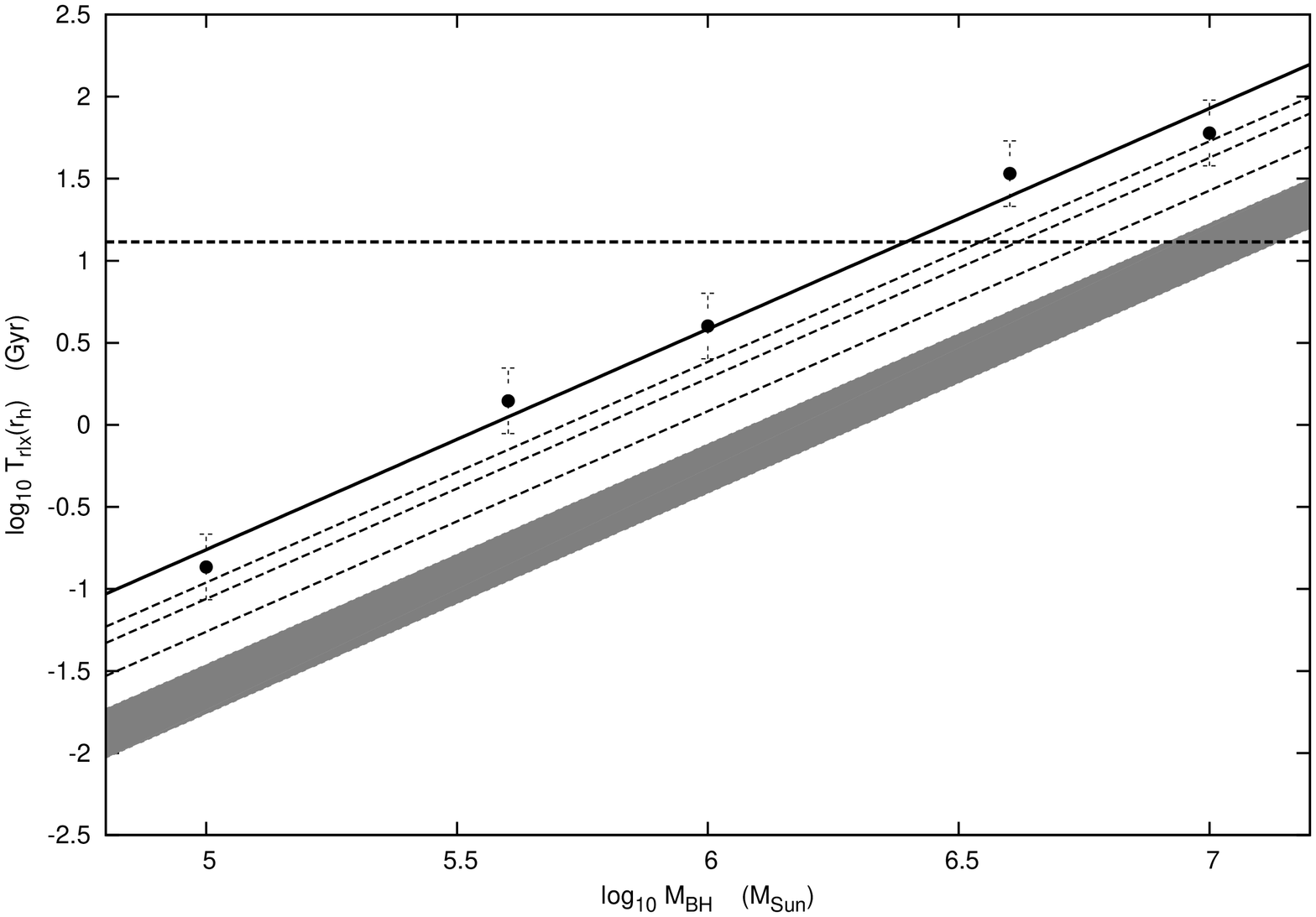}}}}
\vspace{-4pt}
\noindent
$[0.1 T_{\rm rlx},0.2T_{\rm rlx}]$---the time for cusp re-growth if there is no hole in
the initial DF. The three dashed lines above the shaded region represent the
average time needed for the cusp re-growth in case one imposes an initial cavity
with size $R_{\rm cav}=0.5, 1$ or $2$ pc. The horizontal dashed curve represents
13 Gyr. It can be seen that the time needed to re-grow a cusp around MBH with
masses $M_\bullet \lesssim 5 \times 10^6 M_\odot$ is below a Hubble time so
long as the initial cavity is smaller than $\lesssim 2$ pc.
\vspace{8pt}

Nevertheless, it has come as a surprise that very recent spectroscopic observations 
of the Galactic center revealed a core (or even a dip) in the surface distribution 
of the old stellar population (essentially red giants) which should have had time to 
relax into a cuspy density profile \citep{2009A&A...499..483B,2009ApJ...703.1323D}.
The caveat is that the detected stars are still a small fraction, of about 5\%, of 
the stellar population as a whole and therefore do not exclude the presence of an 
extended dark cluster (presumably made of stellar bhs and other CRs)---which would 
indeed agree with our theoretical expectations.

\subsection{Carving a hole in the stellar distribution}
To assess the likelihood that the Galactic center is indeed unrelaxed, it is natural 
to ask: how long does it take to re-grow a cusp if, at some point, it has been destroyed? 
A complete answer will, of course, depend on the extent to which the cusp was destroyed, 
{\it i.e.} how much mass was expelled from the original cusp and over which radial range. 
At this level, it does not matter very much which mechanism led to the destruction of the 
cusp. We discuss briefly possible scenarios for cusp destruction at the end of the section.

In order to investigate this question, we have concocted a set of initial conditions purported 
to mimic the outcome of a destroying cusp event---such as the carving of a cavity in phase 
space through the ejection of stars by, say, an infalling IMBH or, following a major merger, by 
a MBH. We model the outcome of such an event by imposing that all stars with binding energies 
larger than some $E_0$ or almost equivalently, with semimajor axis smaller than $GM_\bullet/2E_0$, 
are not present in the initial DF. In fact, inside $r_h$ the MBH dominates the gravitational 
potential and $E \sim G M_\bullet/2a$. We thus set up an initial Dehnen model with $f(E)=0$ for 
$E>E_0$---in other words, there is an initial cavity in the phase space DF, but not in physical 
space as the stars with lower energy still entail $\rho(r) \sim r^{-1/2}$ at the center, although 
with a smaller amplitude than the original model. The values of $E_0$ were chosen to represent
cavities of size $R_c=0.5, 1, 2$ pc resulting from the inspiral of an IMBH/MBH. Note that these 
models are, by construction, 
isotropic in the velocity distribution. \footnote{We assume that the timescale for isotropization 
of velocities is much shorter than that associated with the cusp re-growth; in any event this 
should not affect our estimates by more than $10\%$ or $20\%$ maximum.} Our fiducial model is a 
Milky Way type nucleus with $M_\bullet = 4\times 10^6 M_\odot$, some $10^6 M_\odot$ in total stellar 
mass inside $1$ pc distributed according to an initial central density slope $\gamma=1/2$, two 
components with masses $m_L=1 M_\odot$ and $m_H=10 M_\odot$, and $0.1\%$ of stellar bhs by number. 
When the stellar distribution has no phase space cavity, this translates into having stars down 
to roughly $10^{-5}$ pc. Having validated the FP models with detailed $N$-body simulations, we
now proceed in the rest of the paper to describe results obtained with the (much faster) FP 
approach.

Figure 2 shows the times for cusp re-growth computed with FP for different galactic nuclei models. 
The time for cusp re-growth is defined as the time it takes for a given initial density profile 
($\rho(r)$ in space or $f(E)$ in phase space, with or without an initial cavity) to reach its
asymptotic slope, which depends on $\Delta$, down to $r \sim 0.01 r_h$. This is indeed the scale
which is resolved by recent observations of the Galactic center \citep{2009A26A...502...91S}.
The shaded region represents the time of cusp re-growth for a range of $R$ 
and $f_H$ (all in the SMS regime, $\Delta<1$) for the case where $f(E)$ extends to high $E$ without 
any cut. It can be seen that, for $M_\bullet \lesssim 10^7 M_\odot$, cusps grow in less than a Hubble 
time; in the particular case of the Milky Way nucleus with $M_\bullet \sim 4 \times 10^6 M_\odot$, it 
takes no longer than $\sim 4.8$ Gyr to fully re-grow a steady-state, mass segregated, stellar cusp 
and only $\sim 2.4$ Gyr to have it grown down to $0.01 r_h$.  If, instead, an initial cavity is imposed 
at the center with size $R_{\rm cav}=0.5, 1$ or $2$ pc in case of the Milky Way (or $R_{\rm cav}=0.2 r_h, 
0.4 r_h$ or $0.8 r_h$ in case of a generic nucleus), times for re-growth are represented by the 
dashed curves above the shaded region. In this case, times for cusp re-growth increase; in the Milky 
Way case, it becomes $\sim 4.8, 7.2$ or $12$ Gyr, respectively. Note that, in the mass range 
$10^5-10^6 M_\odot$, the times for cusp re-growth are definitely {\it much shorter} than a Hubble 
time---even if a fairly large cavity of size comparablee to $r_h$ is hypothesized. The full curve 
represents the relaxation times computed at the radius of influence $r_h$, while the dashed curves 
represent the {\it actual} times for cusp re-growth as measured from FP calculations \citep{PA10}. 

It is difficult to devise plausible mechanisms for the formation of such large cores in the stellar
distribution. For instance, the inspiral of an IMBH of mass $M_\bullet \sim 10^{3-4} M_\odot$ that forms 
an unequal-mass binary with the MBH and ejects stars through three body encounters would tend to 
progressively wipe out the stellar cusp. However, the core radius carved by such an event is $r_{c} 
\sim 0.02-0.04$ pc \citep{2006MNRAS.372..174B} and thus a steady inflow of such IMBHs (one every $10^7$ 
years for a Hubble time) would be required in order to carve a large core $50$ or $100$ larger. Such 
large inflow of IMBHs have been proposed by \cite{2006ApJ...641..319P}. This does not seem very likely 
anymore in light of the fact that such IMBHs were hypothesized to be formed by runaway mergers of stars 
in the center of globular clusters. However, at solar metallicities, such mechanism seems very inneficient. 
Mass loss due to very strong winds severely limits the growth of the stellar object being formed and the 
likely end result of a runaway merger is a $\sim 100 M_\odot$ Wolf-Rayet star. At lower metallicities, 
mass loss is lower and the remnant can be more massive $\sim 260 M_\odot$, but in any case it will not 
form an IMBH \citep{2009A&A...497..255G}. In sum, it looks very unlikely that sufficient IMBHs can be 
formed in order to generate such steady inflow to the Galactic center. Another possibility would be 
that SgrA* is a binary MBH, but this would most likely imply that there has been a, more or less recent, 
major merger involving the Milky Way. This would contradict the apparent pure-disk nature of 
the Galaxy, as theoretical interpretations of stellar kinematic data of the Galactic Bulge seem to 
favor that the Bulge is part of the disk and not a separate component resulting from a merger
\cite{2010ApJ...720L..72S}y---aside from the fact that there are strong constraints from the 
SgrA* proper motion \citep{2004ApJ...616..872R}.

Stars in a Keplerian potential, $G M_\bullet/r$, do not precess because of the $1:1$ resonance 
between their radial and azimuthal frequencies. Resonant relaxation (RR) results from the coherent 
torques that such stars exert on each other leading to a fast evolution of their orbital angular 
momenta over timescales $\lesssim T_{pr}$, where $T_{pr}$ is the precession timescale due to departures 
from an exact Kepler potential \citep{1996NewA....1..149R}. \cite{2010arXiv1010.1535M} suggest that 
RR, by increasing the rate of angular momentum diffusion in the near-Keplerian gravitational potential 
around the MBH, may boost the tidal disruption rate of stars and could thus 
create a near-cavity (out to $\sim 0.1$ pc) in the stellar distribution. It is certain that RR 
operates to some extent in the inner parsec, but we doubt it can completetly explain the dearth 
of red giant stars there or, more generally, the full destruction of a cusp---including CRs such 
as stellar bhs\footnote{Resonant relaxation is, nevertheless, very likely to have a significant 
impact on the EMRI event rates \citep{2006ApJ...645.1152H}.} First, their final density 
distribution does not show a cavity, nor a shallow cusp profile, for $r \gtrsim 0.1$ pc; instead, 
they get final slopes $\gamma \sim 1.5$ for $r \gtrsim 0.1$ pc. This is in contrast with observations 
which show a 
decaying density for $r \lesssim 0.24$ pc \citep{2009A&A...499..483B}. Second, we believe that 
their Monte-Carlo calculations severely underestimate the rate at which the cusp re-grows; in fact, 
they obtain a timescale which is $\sim 10-30$ times longer than that obtained with our $N$-body 
simulations (which are free from any simplifying dynamical assumptions), and also from ours and 
their own recent FP calculations \citep{2010arXiv1002.1220H}. Third, were they to use the latter 
rates, and given that the time RR takes to deplete the cusp is of the same order as the 
time we obtain for cusp re-growth, the net effect of RR on the cusp would likely be minute. 
Moreover, stellar bhs cannot be tidally disrupted, and this makes them less susceptible to be 
extracted from the cusp than $10 M_\odot$ stars.

\section{EMRIs rates}

\subsection{Adiabatic and abrupt EMRIs: Estimation of the rates}

Given a steady state stellar bhs continue to diffuse in $(E,J)$-space and some of them eventually 
come into close interaction with the MBH. During a close interaction, a stellar bh can either be 
promptly scattered into the MBH, accompanied by a single or a few brief bursts of GWs in the 
{\em LISA} band---the so-called ``direct-plunges''---, though they are not likely detectable unless 
if emitted from the Galactic center \citep{2007MNRAS.378..129H}, or scattered outwards in the cusp. 
In either case, it does not live enough to become an EMRI. Alternatively, it may undergo a very slow, 
adiabatic, inspiral without being appreciably disturbed by other stars and, in this case, it will 
eventually become an EMRI detectable by LISA. An EMRI object thus has to spend very many orbits 
without being significantly scattered by the gravitational tugs of the other stars. In other words, 
they must fullfill the following {\it inspiralling criterion}: the time $T_{\rm GW}$ it takes for the 
inspiral, due to orbital energy lost by GW emission only, must be shorter than the typical time $T_J$ 
it takes on average to drift in angular momentum by an amount $J$ which equals its orbital angular 
momentum. Otherwise, they will be promptly captured by the MBH before entering the LISA band. The 
inspiral criterion can be stated in terms of the parameter $s$ being smaller than unity, 
$s=T_{\rm GW}/T_J < 1$. For $T_{\rm GW}>T_J$, it is almost certain that this object has either taken 
an almost radial orbit and fallen into the MBH as a direct plunge or has been scattered outwards. 
\footnote{In steady state, on average each star that drifts outward by an amount $J$ will be 
compensated by another that drifts inward by the same amount. This balance only breaks down for 
those orbits that fall on to the hole, since there are obviously no stars coming out of it to  
keep detailed balancing.} 
It turns out that this parameter simply scales  with  orbital's semimajor axis: $s \propto a^{3/2-p}$ 
\citep{HopmanAlexander05}, which means that it is a decreasing function of $a$ so long as $p<3/2$. 
This is indeed the case in both regimes of mass segregation. Furthermore, \cite{HopmanAlexander05} 
have shown that the probability for a successful inspiral as a function of orbital semimajor axis 
(or energy) is almost a step function of semimajor axis. If $a<a_{\rm GW}$, it is almost certain that 
the stellar bh will become an EMRI; it will almost certainly not become one in case the inequality 
sign is reversed (and the width of the ``transition region'' is very small). This crucial threshold 
quantity demarcates the orbits which are close enough to the MBH to sucessfully decouple from the 
rest of the cluster and undergo the slow, adiabatic inspiral that defines an EMRI from those more 
weakly bound orbits that will be perturbed out of the EMRI tracks due to scattering with other stars.

Therefore, in order to estimate the EMRI event rate, given a steady state $f(E)$, obtained via the FP 
equation, one essentially counts the number of stars that populate the region of phase space for which 
the inspiralling criterion above is satisfied and divide it by the local relaxation time. Note that here, 
for simplicity, we ignore other driving mechanisms---in particular, we ignore resonant relaxation. Under 
these assumptions, the EMRI rate for stellar bhs is approximately given by

\begin{equation}
\Gamma_{\rm EMRI} = f_\bullet \int_{E_{\rm GW}}^{+\infty} dE\  \frac{n(E)}{\ln(J_c(E)/J_{\rm lc}) \ T_{\rm rlx}(E)},
\label{emri-rate}
\end{equation}
where $f_\bullet$ is the number fraction of SBHs in the stellar population,
$n(E)$ is the total number of stars per unit energy($n(E) \propto f(E)$, see \cite{2010GWN.....3....3P}), 
$J_c(E)=\sqrt{G M_\bullet}/2E$ is the specific angular momentum of a circular orbit of energy $E$, 
$J_{\rm lc}=4\,GM_\bullet/c$ is the loss-cone angular momentum and $T_{\rm rlx}=0.34 \ \sigma^3/[G^2(m_\bullet 
\rho_\bullet +m_*\rho_*) \ln \Lambda]$ is the relaxation time.  The log term in the denominator 
in~(\ref{emri-rate}) arises from the phase space (partial) depletion resulting from the presence of the 
loss cone. The conversion between $r$ and $E$ is, for $r \ll r_h$, $\langle E(r) \rangle = G M_\bullet/2 r$ 
or $E = G M_\bullet/2 a$. The critical radius $a_{\rm GW}$, or energy $E_{\rm GW}$, for EMRIs is approximately 
$a_{\rm GW}=0.01 r_h$; and, to first order, $a_{\rm GW}$ is independent of $M_\bullet$ \citep{HopmanAlexander05}.  

\subsection{The relevance of realistic models of mass segregation for the rates}
The weak regime of SMS, and corresponding BW solution, would lead to a fairly high intrinsic rate, per
galaxy, of EMRIs. In fact, Figure 3 shows that, for a Milky Way nucleus, in case $\Delta > 1$, the 
intrinsic EMRI rate is $\gtrsim 10^3$ per Gyr. This is, however, unrealistic as such scenario pressuposes 
an unrealistically high number fraction of bhs ($f_\bullet \gtrsim 0.0325$ for $\Delta>1$). 
In the more realistic case, when $\Delta \sim 0.03$ the BW solution would entail a strong supression of the 
EMRI rate to---at best---a few tens of events per Gyr. This is where SMS solution appears to rescue us. SMS 
implies a higher density of bhs  inside $r_h$ as compared with the $\gamma=7/4$ solution, and in this way---by 
decreasing the local $T_{rlx}$ and increasing $n(E)$ close to the MBH---it partially, but not 
completely, compensates for the small number fraction of bhs entailed by realistic mass functions.

\vspace{8pt}
{\bf Figure 3.} EMRI rate as a function of $\Delta$. The number of stellar bh EMRI events per Gyr in a 
Milky Way type nucleus ($M_\bullet = 4\times 10^6 M_\odot$ and $M_*(< 1 \rm{pc})=10^6 M_\odot$)
\parpic[r][r]{\mbox{\resizebox{0.6\hsize}{!}
        {\includegraphics[angle=270]{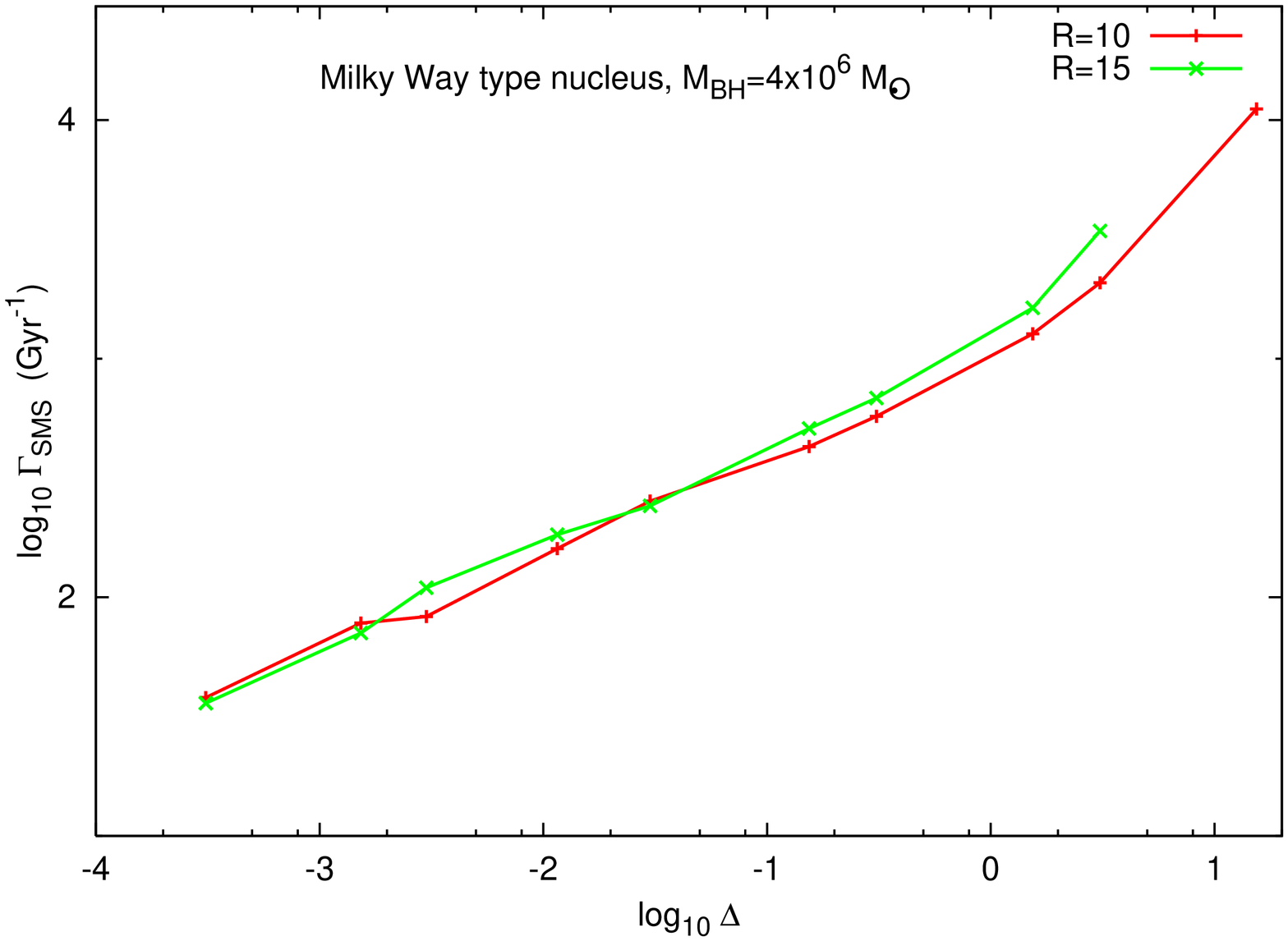}}}}
\vspace{-4pt}
\noindent
as a function of the parameter $\Delta$. This is computed from a two-component
mass segregated stellar cusp ($\gamma_H \approx 2.1$ and $\gamma_L \approx 1.5$) with mass ratios
$R=10$ and $15$ obtained from FP calculations. In the case of the fiducial value $f_\bullet=10^{-3}$, 
$\Delta \approx 0.03$; in those circumstances, each Milky Way like nucleus will produce on average 
$\sim 250$ stellar bh EMRIs per Gyr.
\vspace{8pt}

In order to quantitatively evaluate the boost $\Gamma_{\rm SMS}/\Gamma_{\rm BW}$ to the EMRI rates
from SMS, for a given $\Delta$ and a fixed mass normalization at $r_h$, one needs to estimate what would be 
the rate if the spatial and phase space densities were determined by the $\gamma=7/4$ 
cusp for $r \lesssim 0.1 r_h$. This is done as follows: we define analytically both $\rho(r)$ and $f(E)$ that 
would result from a $\gamma=7/4$ inside $0.1 r_h$
\begin{eqnarray}
\rho(r) & = & \rho_{FP}(r),  \ \ \ \ \ \ \ \ \ \ \ \ r > r_L     \nonumber   \\
\rho(r) & = & \rho_{FP}(r_L) \times  \left( \frac{r_L}{r} \right)^{7/4}, \ \ \ \ \  r \leq r_L ,
\end{eqnarray}
and
\begin{eqnarray}
f(E) & = & f_{FP}(E),  \ \ \ \ \ \ \ \ \ \ \ \ E < E_L     \nonumber   \\
f(E) & = & f_{FP}(E_L) \times  \left( \frac{E}{E_L} \right)^{1/4}, \ \ \ \ \  E \geq E_L ,
\end{eqnarray}
\noindent
where the indices FP mean that the profile is taken from the Fokker-Planck calculation. $r_L$ (and $E_L)$
is a reference radius (energy) chosen according to $r_L \sim 0.1 r_h$.


{\bf Figure 4.} Boost on the EMRI rates from strong segregation. 
\parpic[r][r]{\mbox{\resizebox{0.6\hsize}{!}
        {\includegraphics[angle=270]{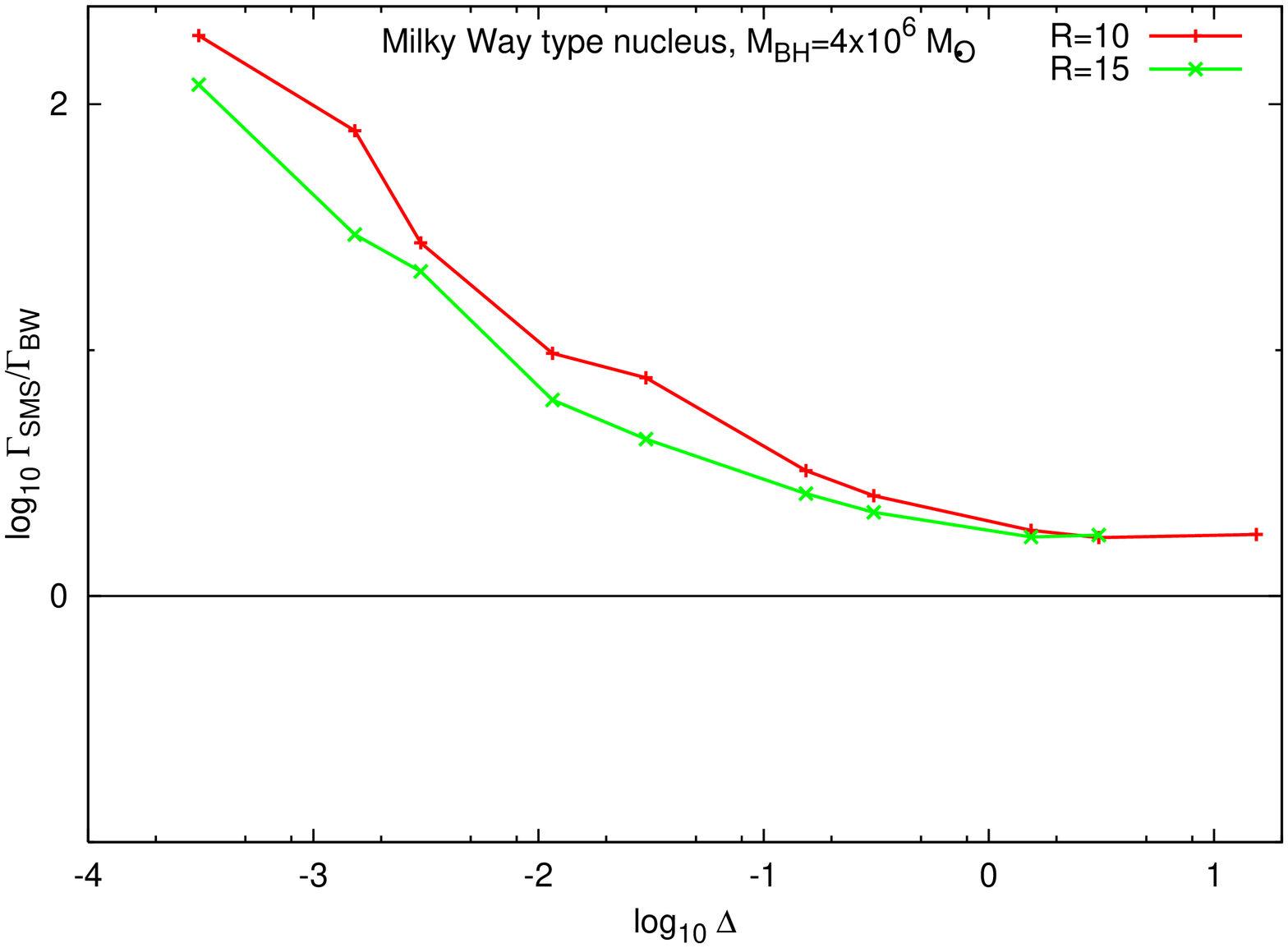}}}}
\vspace{-4pt}
\noindent
One can see that, for values of $\Delta < 1$, there is a significant boost to the EMRI rates in
comparison to which it would result in the case of a $\gamma=7/4$ BW cusp. In particular, for our 
fiducial value $\Delta \sim 0.03$ ($f_\bullet \sim 10^{-3}$), the boost is of order of a factor $10$ 
with respect to a $7/4$-BW cusp with the same mass normalization at $r=1$ pc.
\vspace{8pt}

Figure 4 shows the boost to the EMRI rates due to SMS relative to what
would be obtained from a BW profile. Going from an unrealistically high
$f_\bullet$, as adopted by BW77 (say $\Delta=3$), to a more realistic $f_\bullet$
(say $\Delta=0.03$), while neglecting the existence of SMS, one supresses the
EMRI rate by factors of $\sim 100-150$  (the former would lead to $\sim
\rm{few} \times 10^3$ EMRIs per Gyr; the latter is reduced to $\sim $ few tens
per Gyr). However, by taking into account the SMS solution, for this low
$\Delta=0.03$, we boost again the rates by a factor close to $10$, thus partially
compensating the reduction of EMRIs (from few tens to a few hundred per Gyr; in
fact, there are  $\sim 250$ per Gyr in case $\Delta=0.03$ for a Milky Way nucleus). 
We conclude that the apparently inocuous and tiny change of the logarithmic slope 
from $\gamma_H=7/4$ to $\gamma_H \sim 2$ can have a substantial effect (a factor 
of $\sim 10$) on the expected EMRI rate. 

{\bf Figure 5.} EMRI rates as a function of MBH mass in strongly segregated
nuclei. 
\parpic[r][r]{\mbox{\resizebox{0.6\hsize}{!}
        {\includegraphics[angle=270]{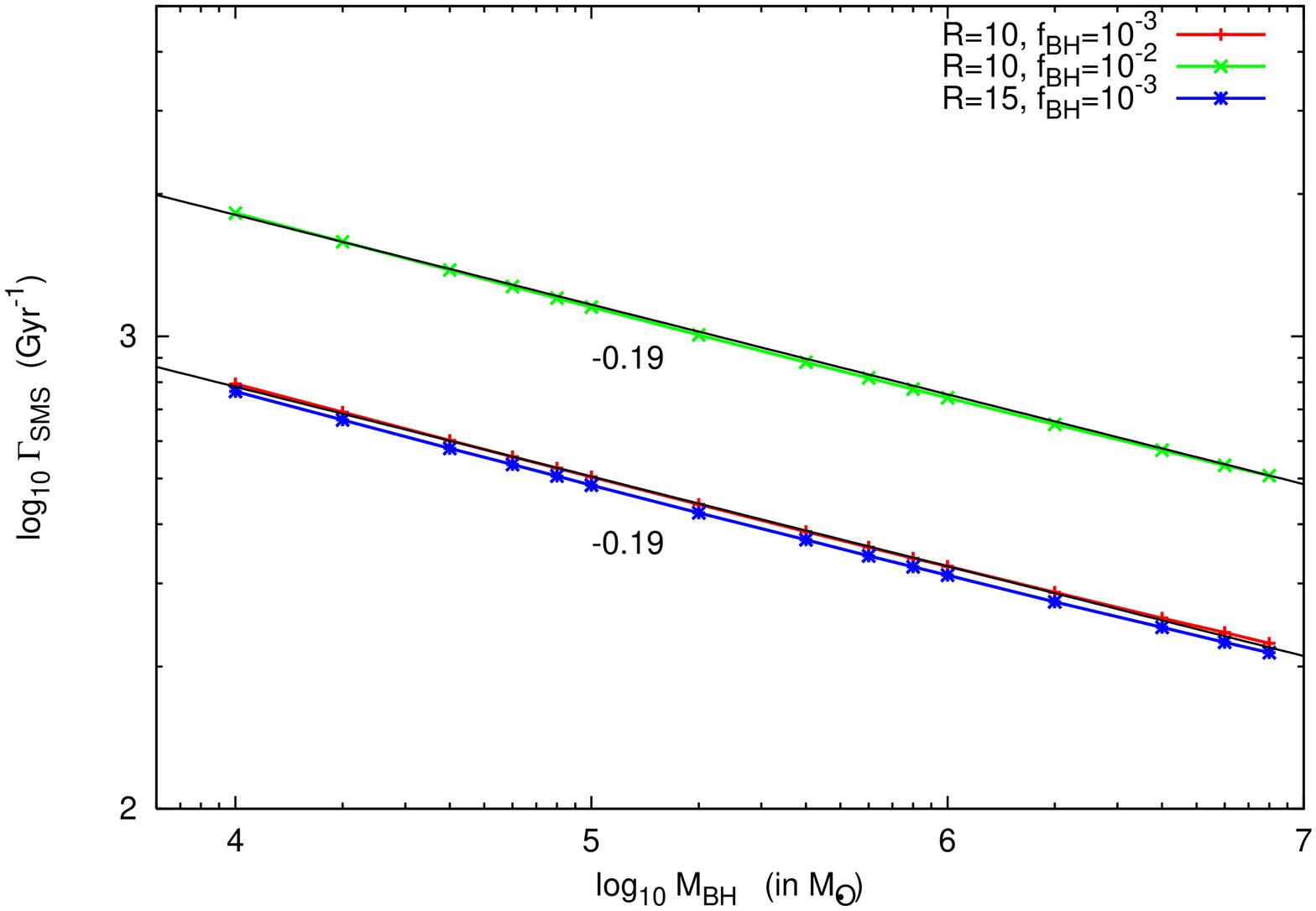}}}}
\vspace{-4pt}
\noindent
The EMRI rate depends on the MBH mass, $\Gamma_{SMS} \propto M_\bullet^{-0.19}$. Shown are
curves for $\Delta=0.03$ ($f_\bullet=10^{-3}$ and $10^{-2}$) and three different mass ratios 
between heavy and light stars, $R=10$ and $15$. 
\vspace{8pt}

Figure 5 shows the dependence of the intrinsic EMRI rate on the mass of the central
MBH, where the validity of the $M_\bullet-\sigma$ relation was assumed \citep{2005SSRv..116..523F}. 
We see that the EMRI rate for stellar bhs scales as $\propto M_\bullet^{-0.19}$, independently
of $R$ and $f_\bullet$. Its absolute normalization depends obviously on the number fraction
$f_\bullet$ of sbhs, in agreement with Figures $3$ and $4$.

One can make a rough conversion of the estimated intrinsic rate into LISA detection rates.
Following \cite{Gair09}, who made a number of assumptions regarding the local density of
MBHs and its spin distribution, plus on the LISA detection capabilities, we find that 
according to its equation ($7$), LISA will see around $\sim 10^2-7 \times 10^2$ EMRI
events during a $2$-year or $5$-year mission, respectively. Note that these rates may
change by factors of $\sim 2-3$ as a function of corrections to the local MBH mass density
\citep{Graham-Driver07}; moreover, the uncertainties regarding the efficiency of RR and
other channels may still affect the rate predictions by one or two orders of magnitude.
A lot of work still remains to be done; nevertheless, the consequences regarding the
SMS regime are significant and under control.

\section{Conclusions}

We have considered simplified stellar models of galactic nuclei, with only two mass 
components, which harbor MBHs that fall into the LISA detection bandwidth. For quite 
generic initial conditions, such stellar clusters are expected to have reached a relaxed, 
mass segregated, steady state which is independent of initial conditions at the time 
of formation. Strong (realistic) mass segregation is a robust outcome from the growth 
and evolution of stellar cusps around MBHs in the mass range $10^4-10^7 M_\odot$ to which 
LISA will be sensitive. Our N-body results validate the Fokker-Planck description of the 
{\it bulk} properties of the stellar distribution. SMS boosts the EMRI event rates with 
respect to what would be implied by a shallower stellar density profile ({\it e.g.} 
$\gamma=7/4$, which has been the working assumption of almost all event rate estimates 
in the literature so far) that also respect the mass normalization obtained from 
observations of the Galactic center at $1$ pc from the hole. In particular, our fiducial 
models of the Galactic center are enhanced by a factor of $\sim 10$---leading to a predicted 
value of $\sim 250$ stellar bh EMRIs per Gyr. The FP formalism assumes two-body relaxation 
as the only dynamical driver present---this could be a severe restriction at radii $\lesssim 
0.01 r_h$, inside which even the NB simulations with higher $N$ in our sample start to run 
out of particles and where RR could play an important role \citep{2006ApJ...645.1152H,
2010arXiv1010.1535M}. Other crucial mechanisms are resonant relaxation, (small) triaxiality 
of the galactic potential, tidal separation of binaries and massive perturbers \citep[see
e.g.][ for a review]{Amaro-SeoaneEtAl07}.  These are the subject of our current
research work, and the extent to which they can significantly affect the EMRI rates is
still an open question.

\ack

It is a pleasure for PAS to thank Tom Prince for the invitation to give a
plenary talk in the symposium. He is equally indebted with Esmeralda
Mart{\'\i}nez for discussions as well as with Vivian J. Drew for her great
sense of humour and coordination of the workshop. MP thanks Lauren B. for
interesting comments. PAS and MP are partially
supported by DLR (Deustches Zentrum f\"ur Luft- und Raumfahrt). The simulations
have been carried out on the dedicated high-performance GRAPE-6A clusters at
the Astronomisches Rechen-Institut in Heidelberg \footnote{GRACE: see
http://www.ari.uni-heidelberg.de/grace}, which was funded by project
I/80\,041-043 of the Volkswagen Foundation and by the Ministry of Science,
Research and the Arts of Baden-W\"urttemberg (Az: 823.219-439/30 and /36), an
also at the {\sc Tuffstein} cluster of the AEI.

\end{document}